\documentclass[apj,twocolumn]{emulateapj}
\usepackage{epsfig,apjfonts,mathptmx}

\shorttitle{Metallicity dependence of the IMF}
\shortauthors{Zhang et al.}

\newcommand{\gsim}{\lower.5ex\hbox{$\; \buildrel > \over \sim \;$}}
\newcommand{\lsim}{\lower.5ex\hbox{$\; \buildrel < \over \sim \;$}}
\newcommand{\ha}{\hbox{H$\alpha$}}
\newcommand{\hb}{\hbox{H$\beta$}}
\newcommand{\hg}{\hbox{H$\gamma$}}
\newcommand{\hd}{\hbox{H$\delta$}}
\newcommand{\he}{\hbox{H$\epsilon$}}
\newcommand{\oii}{\hbox{[O\,{\sc II}]}}
\newcommand{\nii}{\hbox{[N\,{\sc II}]}}
\newcommand{\sii}{\hbox{[S\,{\sc II}]}}
\newcommand{\oiii}{\hbox{[O\,{\sc III}]}}
\newcommand{\neiii}{\hbox{[Ne\,{\sc III}]}}
\newcommand{\feiii}{\hbox{[Fe\,{\sc III}]}}
\newcommand{\ariv}{\hbox{[Ar\,{\sc IV}]}}

\newcommand{\hii}{\hbox{H\,{\sc II}}}
\newcommand{\heii}{\hbox{He\,{\sc II}}}
\newcommand{\hei}{\hbox{He\,{\sc I}}}
\newcommand{\caii}{\hbox{Ca\,{\sc II}}}
\newcommand{\zoh}{\hbox{$12\,+\,{\rm log(O/H)}$}}
\newcommand{\etal}{et al.\ }

\begin{document}

\title{Wolf-Rayet Galaxies in the Sloan Digital Sky Survey:
 the metallicity dependence of the initial mass function}

\author{
Wei Zhang\altaffilmark{1}, 
Xu Kong\altaffilmark{1}, 
Cheng Li\altaffilmark{1}, 
Hong-Yan Zhou\altaffilmark{1}, 
Fu-Zhen Cheng\altaffilmark{1}}

\altaffiltext{1}{Center for Astrophysics, University of Science and
Technology of China, 230026, P. R. China}

\begin{abstract}
We use a large sample of 174 Wolf-Rayet (WR) galaxies drawn from the 
Sloan Digital Sky Survey to study whether and how the slope of the 
stellar initial mass function depends on metallicity. 
We calculate for each object its oxygen abundance according to which 
we divide our sample into four metallicity subsamples. 
For each subsample, we then measure three quantities: the equivalent 
width of \hb\ emission line, the equivalent width of WR bump around 
4650\AA, and the WR bump-to-\hb\ intensity ratio, and compare to the 
predictions of the same quantities by evolutionary synthesis models 
of Schaerer \& Vacca. 
Such comparisons lead to a clear dependence of the slope of initial 
mass function ($\alpha$) on metallicity in that galaxies at higher 
metallicities tend to have steeper initial mass functions, with the 
slope index ranging from $\alpha\sim$1.00 for the lowest metallicity 
of $Z=0.001$ to $\alpha\sim$3.30 for the highest metallicity $Z=0.02$. 
We have carefully examined the possible sources of systematic error 
either in models or in our observational measurements and shown that 
these sources do not change this result. 
\end{abstract}

\keywords{galaxies: abundances -- galaxies: starburst -- stars:
	 Wolf-Rayet }

\section{Introduction}

The stellar initial mass function (IMF) describes the relative
frequency with which stars of various masses are formed, and so
controls nearly all aspects of the evolution of stellar systems. 
An understanding of the IMF has thus long been one of the major 
goals in many astronomical fields, from theories of star formation 
to the interpretation of integrated properties of galaxies at the 
highest redshift.

The studies of IMF began with the publication of the first and still 
most famous paper on the subject by Salpeter (1955), who found that 
the solar neighborhood IMF can be approximated by a declining power 
law, $dN/dM \propto M^{-\alpha}$, with a slope of $\alpha=2.35$. From 
then on, there have been considerable observational studies aiming to 
determine the form of the IMF (e.g. Scalo 1986, 1998; Kroupa 2001, 
2002; Reid \etal 2002; Chabrier 2003). 
These studies have established that, for masses above 1 $M_\odot$, the 
IMF can generally be quantified using the Salpeter form, while for 
low masses it deviates from this form, by flattening below 
0.5 $M_\odot$, peaking at a characteristic mass of $\sim 0.1 M_\odot$ 
and then significantly declining. However, despite much
progress, there are still many problems that are far from being
solved (Kennicutt 1998; Bate \& Bonnell 2005), out of which the
universality or the variation of the IMF is among the most fundamental 
and is still a controversial matter. In particular, it is not very 
clear whether or not the IMF, including both its characteristic
mass around $0.1 M_\odot$ and its power-law slope at higher masses,
depends on environmental conditions such as metallicity.

One reliable way to derive an IMF is to count individual stars when 
both photometry and spectroscopy are available. Massey \etal have 
applied this method to numerous young star clusters and OB 
associations in the Galaxy and the Magellanic Clouds (Massey \etal 
1995, Massey 1999).
This approach is, however, infeasible for deriving the starburst IMF
because the distance of the closest starburst galaxies are $\gsim$
2Mpc, where the spectrum of individual stars cannot be 
obtained. Thus the stellar content of galaxies at large distances
can only be derived from their integrated spectra
(Leitherer 1998). In such studies, many authors have preferred to
Wolf-Rayet (WR) galaxies, where broad emission lines from WR stars are
observed in the integrated spectra. WR galaxies have long been known,
beginning with the discovery of such features in the spectrum of the
galaxy He 2-10 by Allen, Wright, \& Goss (1976). Osterbrock \& Cohen
(1982), and Conti (1991) introduced the concept of WR galaxies. These
galaxies are thought to be undergoing present or very recent star
formation that produces massive stars evolving to the WR stage, and so 
are ideal objects for studying the early phases of starbursts,
determining burst properties and constraining parameters of the IMF
(Schaerer \etal 1999; Guseva \etal 2000).

Many studies have been done trying to constrain the main burst
parameters, such as the age and duration of the bursts and the IMF, 
by comparing the observed WR features with that predicted by 
population synthesis models for young starbursts (Contini \etal 1999; 
Schaerer \& Vacca 1998, hereafter SV98). These studies have produced 
some observational evidence in support of the variation of the IMF 
slope with metallicity. Schaerer \etal (1999) reported the detection 
of WC stars in five WR galaxies, of which three have metallicities of 
$Z=0.004$ and the rest two have $Z=0.008$ and $Z\gsim0.020$ 
respectively. By
comparing the observations with the model predictions of SV98, these
authors found that only the two objects with higher metallicities are
consistent with the standard model in which a Salpeter IMF is assumed,
whereas the observed WR emission lines of the other three low 
metallicity objects can not be reproduced by this model and a flatter 
IMF is required in order to properly bring the models into agreement 
with the data. 
A study of three WR galaxies with $Z=0.004$ by Huang \etal (1999) 
reached a similar conclusion that a top-heavy IMF with $\alpha = 1.00$ 
is likely required to favor the formation of WR stars in these 
low-metallicity galaxies. Recently, Schaerer \etal (2000) investigated 
six metal-rich WR galaxies ($Z\gsim$0.020) and found that a very steep 
IMF with $\alpha\gsim3.30$ is very unlikely for these objects. 
A more recent analysis of a larger sample of 14 WR galaxies by Fernandes 
\etal (2004) has concluded that there does exist some dependence of the 
IMF slope on galaxy metallicity. In this study, low-metallicity galaxies 
are predicted to undergo a burst of star formation and show a flat IMF 
varying between the top-heavy and the Salpeter form, whereas high 
metallicity galaxies either show a steeper IMF or experience an extended 
burst.

The studies described above suggest that, at least for WR galaxies, the
slope of the IMF is somewhat dependent on metallicity. However, the
galaxy samples used in these previous studies are not sufficiently large 
to give reliable constraints on the relation between the IMF and the
metallicity of galaxies. In this paper, we address this problem using a
large sample of WR galaxies drawn from the Data Release Three (DR3;
Abazajian \etal 2005) of the Sloan Digital Sky Survey (SDSS; York et
al. 2000). SDSS is to date the most ambitious imaging and spectroscopic
survey, providing a library of hundreds of thousands of galaxy spectra
with a resolution of $\la$ 3\AA\ at 5000\AA\ and a median
signal-to-noise ratio (S/N) of $\sim 14$ per pixel. It thus has allowed
one to select an unprecedented large, high quality spectral sample of WR
galaxies, and subsequently to give reliable constrains on the relation
between galaxy metallicity and the IMF slope.

From DR3, we select a large sample of 174 galaxies with obvious WR
features using a two-step method, in which we first select star-forming
galaxies with evident \he\ emission and then select WR galaxies from 
these candidates by visually examining their spectra (\S2). 
Next, for each WR galaxy we carefully calculate its oxygen abundance and 
measure its WR and \hb\ emission, as quantified by the equivalent width (EW) 
of the WR bump around 4650\AA, the EW of \hb\ emission line, and the 
WRbump/\hb\ intensity ratio (\S3). We then investigate the metallicity
dependence of the IMF slope by comparing these observed quantities
with those predicted by evolutionary synthesis models (\S4). We discuss 
in \S5 the possible biases in our results caused by a few sources of
systematic error. Finally, we summarize our results in the last section.

\section{Sample Selection}	

\subsection{The SDSS spectroscopic sample}

The data analyzed in this study are drawn from the SDSS. The survey
goals are to obtain photometry of one-quarter of the sky and spectra
of nearly one million objects. Imaging is obtained in the {\it u, g,
r, i, z} bands (Fukugita \etal 1996; Smith \etal 2002) with a special 
purpose drift scan camera (Gunn \etal 1998)
mounted on the SDSS 2.5-meter telescope at Apache Point Observatory.
The imaging data are photometrically (Hogg \etal 2001) and
astrometrically (Pier \etal 2003) calibrated, and used to select
stars, galaxies, and quasars for follow-up fibre spectroscopy.
Spectroscopic fibres are assigned to objects on the sky using an
efficient tiling algorithm designed to optimize completeness (Blanton
\etal 2003). The details of the survey strategy can be found in York
et al. (2000), and an overview of the data pipelines and products is
provided in the Early Data Release paper (Stoughton et al. 2002).
More details on the photometric pipeline can be found in Lupton \etal
(2001).

Our parent sample for this study is composed of 374,767 objects which
have been spectroscopically confirmed as galaxies and have data
publicly available in the SDSS DR3. It is a catalogue of local
galaxies (mostly below $z\approx 0.3$ and a median $z$ of 0.1),
covering 4188 deg$^2$ on the sky. The spectra are obtained with two
320-fibre spectrographs that are also mounted on the SDSS 2.5-meter
telescope. Fibers with 3 arcsec in diameter are manually plugged into
custom-drilled aluminum plates mounted at the focal plane of the
telescope. The spectra are exposed for 45 minutes or until a fiducial
S/N is reached. The median S/N per pixel for galaxies in the main
sample is $\sim 14$. The spectra cover an optical wavelength range
from 3800 to 9200 \AA\, with an instrumental resolution of $R=1850 -
2200$ (FWHM $\sim 2.4$ \AA\ at 5000 \AA\ ).

\subsection{Description of the selection method}

We aim to select a sample of galaxies showing evident WR features in
their spectra (blue bump around 4650\AA\ and/or red bump around
5808\AA). An intuitively accessible way is to systematically measure 
the EW of WR bumps and take those with large EWs as WR galaxies. 
However, this method will be extremely unefficient when dealing with 
large data sets, given that the WR galaxy population is just a very
small fraction of the whole sample and in most cases systematical
measurements of WR bump strength suffer from large uncertainties in
the continuum determinations and/or from the contamination by the
emission of non-WR stars.

To avoid this difficulty, we instead opt for a two-step method in
which we first select star-forming galaxies with evident \he\ emission 
line as candidates, and then select WR galaxies by visually
examining these candidates. There is no denying that this method will
inevitably drop a number of WR galaxies, especially those with low
S/N. However, this method can still work well for our purpose,
because of the following two facts. First, as described in the
introduction, WR galaxies are believed to be undergoing present or very
recent star formation that produces massive stars. It is then natural to
speculate that a sample of WR galaxies must be a subsample of
star-forming galaxies and most WR galaxies must show high-order Balmer
emission lines in their spectra because these lines are usually 
associated with massive stars that are expected to be numerous in WR 
galaxies. We use the \he\ emission line rather than other Balmer lines 
such as \hd\ in order to ensure that the high-excitation \hii\ regions 
are selected. It should be pointed out that the \he\ emission line is 
in fact a blend of three lines, including \he, \neiii\ and \hei\ lines. 
The latter two lines are often strong in active galactic nuclei (AGN). 
By using this blend we can find AGN with WR features in a harmonious way 
as normal galaxies, which are also interesting and will be discussed in 
a separated paper, at the cost of somewhat reducing our selection 
efficiency for the purpose of this paper. The second fact is that the 
parent sample is substantially large, allowing one to select a large 
enough sample of WR galaxies and thus to study their statistical 
properties still with unprecedented accuracy, even though some objects 
must have been thrown away.

\subsection{Starlight removal and emission line measurements}

To select star-forming galaxies, we have performed a careful
subtraction of the stellar absorption-line spectrum before measuring 
the nebular emission lines. Our methodology for modelling and
subtracting the underlying starlight has been described in detail in 
Li \etal (2005) and in Lu \etal (2006). The method for measuring the
emission-line parameters has been described in Dong \etal (2005).
A set of six absorption-line templates with zero velocity
dispersion are constructed in the first place, by applying the
technique of Independent Component Analysis to the synthesis
galaxy spectra from a newly released high-resolution evolutionary
model by Bruzual \& Charlot (2003). 
The physical meaning of the six templates could be easily 
understood through examining their spectra visually (see
Fig.4 of Lu \etal 2006). The first template represents the blue
continuum of O stars as well as the \caii\,H and \caii\,K lines at 
3933\AA\ and 3968 \AA. The second template is similar to the
spectrum of B stars, while the absorption lines of neutral metals and
molecules such as TiO are also identified. The third template shows
extremely strong Balmer absorption lines, a Balmer jump at the blue
end, and the \caii\ triplet and more lines of neutral metals and
molecules at the red end. The fourth and fifth templates are somewhat
like hybrids of F-K stars with stronger neutral metal and molecule
lines. The last template is similar to the spectrum of M stars in the
long-wavelength range but shows high-order Balmer absorption lines
at short wavelengths. The spectral properties of the templates imply
a tight correlation between the stellar population age and the
templates (see Lu \etal 2006 for a detailed discussion).
The spectra of all galaxies in the parent sample described above are 
then fitted with the six templates with iterative rejection of 
emission lines and bad pixels. During the modeling, the template 
spectra are broadened to velocity dispersion from 0 to 600 kms$^{-1}$ 
using Gaussian kernel, and an average intrinsic
reddening is added to the model by searching a range of color excess 
$E(B-V)$ and assuming an extinction curve of Calzetti \etal (2000). 
For each spectrum, the best-fitting model is subtracted from the 
original spectrum, yielding a pure emission-line spectrum from which 
the emission-line parameters (the central wavelength, line intensity 
and line width) are measured. 

As discussed in Li \etal (2005), our method of modeling and subtracting 
the underlying stellar component have several important advantages, and 
so the nebular emission lines can be accurately measured from the
starlight-subtracted spectra. First, the templates are derived from a
large sample with full coverage of spectral type.
This ensures a very close match to the true underlying stellar population
of different types of galaxies, including emission-line galaxies that
contain much young stellar components. Second, the templates are not
obtained directly from the observed spectra, but from the modeled
one with zero velocity dispersion, and thus could well match the
absorption-line profiles. Finally, all the emission lines with flux above
$3\sigma$ are carefully masked during the modeling. This is vital to the
nebular line measurements, especially for star forming galaxies. For each
galaxy, the modeled spectrum is subtracted from the observed one, and the
emission lines are fitted with Gaussian functions (see Dong \etal 2005 for
details). Because emission lines and absorption lines are coupled, this
procedure is performed iteratively.

\subsection{Selecting WR galaxies}

Based on the emission-line measurements, star-forming galaxies are
selected from the subset of galaxies with the integrated flux
intensity $F>2 \times10^{-17} erg s^{-1}cm^{-2}$ in the four emission
lines \oiii$\lambda$5007, \hb, \nii$\lambda$6583 and \ha. Following 
Kauffmann \etal (2003), a galaxy is defined to be a star-forming galaxy 
if
\begin{equation}
\log(\oiii/\hb) < 0.61/(\log(\nii/\ha) -0.05) + 1.3.
\end{equation}
This gives rise to a total of 117,275 star-forming galaxies.

The candidates of WR galaxies are then selected from these
star-forming galaxies by requiring EW(\he)$>5\sigma$, where EW(\he) is 
the EW of \he\ emission line and $\sigma$ is its error. As a result, a 
total of 7628 galaxies are selected as our candidates. 
As described above, the spectra have been well fitted and the 
best-fitting stellar component (continuum plus absorption lines) has 
been separated from the emission-line component. Based on these fits,
EW(\he) and $\sigma$ can be measured in an easy and accurate manner. 
When measuring, we have adopted a rest-frame wavelength range of 
3959\AA\ to 3979\AA\ for \he. For each object, we use the best-fitting 
stellar spectrum (see \S2.2), but not the observed spectrum, to 
determine the continuum that will be used for calculating the EW, in 
order to avoid possible contamination by nearby nebular emission lines. 
We have adopted 3905-3915\AA\ for the nearby blue band and 
4010-4030\AA\ for the red. In the computation, the flux fluctuations
provided by the SDSS pipeline have been taken into account.

Finally, we visually examine the spectra of the 7628 candidates and 
keep those with obvious WR features. This gives rise to a sample of 
174 WR galaxies.
In some cases, the selected WR galaxies show both the blue and red 
bumps, while in other cases only the blue bump is visible. We 
tabulate our sample in Table 1 so that the reader would be able to
recover our results or perform other analyses using this sample. 
For each WR galaxy, we list in the table the spectroscopic ID
(Column 1), the intensity ratio of various emission lines relative 
to \hb\ emission line (Columns 2-18), the EWs of \hb\ emission line 
(Column 20), the WRbump/\hb\ intensity ratio (Column 21), the EWs 
of blue WR bump (Column 22) and the oxygen abundance (Column 23).
Before these quantities being measured, the observed spectra have 
been corrected for the Galactic and the internal reddening (see the 
next section). The observed flux of \hb\ emission line is listed in
Column 19.

\begin{table*}
\caption{Parameters of the WR sample.\tablenotemark{d}}\label{tbl-1}
\tiny
\begin{center}
\begin{tabular*}{1.\textwidth}{@{\extracolsep{-0.22cm}}cccccccccccccccccccccccc}
\tableline
\tableline
MJD-Plate-Fiber& \oii & H10 & H9 & 
H8& 
H7 & \hd & \hg & \oiii & \oiii & \oiii & \ha & \sii & \sii & \nii & \nii &
\oii & \oii & \hb & EW & WR & EW & 12+ & method \\

& 3727  & 3798  & 3836 & 3890  & 3970  & 4100 & 4340  & 4363  & 4959 &
5007  & 6563  & 6717 & 6731  & 6548  & 6583 & 7320  & 7330  & 4861 &
(\hb)  & bump  & 4650 & log(O/H)  & \\
(1)& (2)\tablenotemark{a}& (3) & (4) & (5) & (6) & (7) & (8) & (9) & (10) & (11) &(12) &(13) &
(14) &(15) &(16) & (17) &(18) &(19)\tablenotemark{b} & (20) &(21) &(22) & (23) &(24) \\
\tableline
52251-0751-629 & ...\tablenotemark{c} & ... & 6.1 & 19.6 & 29.9 & 27.8 & 50.4 & 7.6 & 164.0 & 515.3 & 280.8 & 9.1 & 6.2 & 1.4 & 4.1 & 4.2 & 1.9 & 847.3 & 66.7 & 5.1 & 3.0 & 8.01 & Te \\
51900-0390-445 & 191.4 & 2.6 & 5.3 & 17.8 & 28.7 & 24.8 & 47.7 & 3.6 & 151.2 & 476.2 & 285.4 & 17.9 & 13.6 & 7.2 & 20.7 & 2.3 & 1.9 & 1941.1 & 87.8 & 6.8 & 5.1 & 8.27 & Te \\
51900-0390-106 & 206.9 & ... & ... & 101.3 & 9.2 & 21.0 & 43.2 & ... & 12.7 & 39.5 & 286.0 & 37.9 & 30.3 & 35.7 & 102.0 & 1.1 & 0.5 & 274.3 & 18.4 & 22.1 & 3.5 & 9.04 & N2 \\
52233-0753-094 & ... & 4.8 & 7.7 & 19.2 & 22.9 & 25.1 & 46.8 & 2.4 & 132.9 & 418.2 & 287.0 & 14.7 & 10.6 & 4.4 & 12.7 & 2.0 & 1.7 & 2323.5 & 51.7 & 17.3 & 7.9 & 8.33 & Te \\
51817-0418-302 & ... & 4.1 & 6.4 & 18.2 & 26.2 & 25.0 & 46.8 & 5.6 & 154.4 & 485.8 & 282.5 & 15.6 & 11.4 & 3.0 & 8.7 & 2.0 & 1.5 & 2271.0 & 79.4 & 7.1 & 5.1 & 8.05 & Te \\
51913-0394-472 & 93.3 & 3.9 & 6.1 & 16.7 & 28.7 & 24.9 & 45.3 & 11.1 & 172.9 & 545.1 & 278.4 & 8.2 & 6.1 & 1.3 & 3.7 & 1.1 & 1.0 & 723.8 & 86.8 & 2.7 & 2.1 & 7.80 & Te \\
51789-0398-294 & ... & ... & 6.4 & 20.0 & 22.2 & 25.1 & 47.3 & 1.1 & 94.8 & 297.6 & 289.6 & 24.1 & 16.9 & 6.7 & 19.2 & 2.2 & 2.3 & 1462.4 & 76.1 & 16.3 & 11.0 & 8.42 & Te \\
51817-0399-322 & 246.6 & 0.4 & 3.6 & 17.9 & 12.9 & 24.1 & 46.2 & 0.5 & 29.5 & 92.0 & 288.1 & 37.8 & 29.7 & 30.0 & 85.8 & 1.8 & 1.6 & 975.6 & 33.8 & 7.0 & 2.0 & 8.09 & Te \\
51882-0426-314 & ... & ... & 7.2 & 25.0 & 27.5 & 27.1 & 48.5 & 10.0 & 120.3 & 379.5 & ... & 9.3 & 7.2 & 0.9 & 2.6 & 1.4 & 0.7 & 989.8 & 193.1 & 1.4 & 2.4 & 7.51 & Te \\
51900-0427-291 & 225.2 & 0.1 & 2.4 & 16.9 & 12.4 & 22.5 & 43.3 & 1.6 & 64.4 & 201.6 & 285.2 & 24.5 & 20.3 & 20.1 & 57.5 & ... & ... & 209.5 & 30.8 & 24.4 & 6.6 & 8.06 & Te \\
\tableline
\end{tabular*}\\[0.5ex]
\end{center}
\scriptsize
a) Columns 2-18 and 21 are intensity ratios relative to \hb\ emission 
line, with $I(\hb)=100$. Before these quantities being measured, the 
observed spectra have been corrected for the Galactic reddening, the 
internal reddening and the stellar absorption. See the text for 
details.\\
b) The observed flux of \hb\ emission line in unit of 
$\rm 10^{-17}~erg~s^{-1}cm^{-2}$, which has been corrected for the 
Galactic reddening, but not for the internal reddening.\\
c) Dots denote the cases in which \ha\ or 
\oiii$\lambda\lambda4959,5007$ lines are clipped, or the 
corresponding lines are not detected.\\
d) Table 1 is presented in its entirety in the electronic edition of 
the Journal. A portion is shown here for guidance regarding its 
form and content.
\end{table*}

\section{Observational Measurements}

\subsection{Correction of internal extinction}

In the previous section, we have described our method for subtracting
the underlying starlight and for measuring the emission lines. 
Before modelling the stellar spectrum of galaxies,
the observed spectra have already been corrected for the Galactic
reddening, but not for the internal reddening.
The internal reddening can be estimated using the Balmer decrements
of the emission lines from \hii\ regions. We didn't apply this
correction before the modelling, because our fitting method is aimed
to model the stellar component of the galaxies, but not the \hii\
regions where the gas dominates the observed spectrum. 
It has been known that the mean color excess estimated from the
observed Balmer decrements is systematically larger than the average
of the stellar extinction (Charlot \& Fall 2000; Li \etal 2005). 
Therefore, we choose to take into account the stellar reddening
by a single average color excess during the modelling. In this way our
program attempts to fit the observed uncorrected spectra (continuum 
plus absorption lines). After the modelling, we measure emission-line
parameters, correct for intrinsic extinction using Balmer decrements,
and use the corrected parameters for further analysis.
Our procedure
of sample selection described above is not sensitive to this
modification. However, when estimating metal abundance based on
measurements of nebular emission lines, it is necessary to carefully
correct for the internal reddening.

In practice, the internal reddening is usually corrected as a whole
using the effective extinction curve $\tau_{\lambda} =
\tau_V(\lambda/5500)^{-0.7}$, which was introduced by Charlot \& Fall
(2000). The effective $V$-band optical depth $\tau_V$, arising from
attenuation by dust in a galaxy, can be written
\begin{equation}
\tau_V = -\frac{ {\rm ln}[F(\ha)/F(\hb)]-{\rm ln}[I(\ha)/I(\hb)] } {
    [(\lambda_{\rm H\alpha}/5500)^{-0.7} - (\lambda_{\rm
    H\beta}/5500)^{-0.7}] },
\end{equation}
where $F(\ha)/F(\hb)$ is the ratio of the observed H$\alpha$ emission
intensity relative to H$\beta$, and $I(\ha)/I(\hb)$ is the theoretical
value of this ratio. 
It has been known that $I(\ha)/I(\hb)$ does not keep constant, but
varies with electron temperature $T_e$, ranging from $\sim$2.7 at
$T_e=20000$K to $\sim 3.0$ at $T_e=10000$K. In order to take this
variation into account, we determine the electron temperature $T_e$
using the observed intensity ratio
\oiii$\lambda4363$/\oiii$\lambda\lambda4959,5007$, from which we then
calculate $I(\ha)/I(\hb)$ using the method introduced by Storey \&
Hummer (1995). As a result, theoretical \ha-to-\hb\ ratios are 
successfully obtained for 152 WR galaxies whose $T_e$ can be 
determined. 
In general, the observed \ha/\hb\ ratio is larger than the 
theoretical value, because \hb\ emission is more heavily absorbed
than \ha. However, out of the 152 WR galaxies, there are 12
objects showing relatively smaller \ha/\hb\ ratios than theoretical 
values. This could be due to clipped \ha\ lines or observational 
flux fluctuations.
For these galaxies, we instead follow Izotov \etal (1994) to use 
\hg/\hb\ intensity ratios to derive the extinction coefficient 
$C(H\beta)$ and the theoretical values of \hg/\hb. In this way, 8 
objects are corrected for the internal reddening. For the remaining 
4 objects whose observed \hg/\hb\ ratios larger than the theoretical 
values, we do not perform the correction.

For the 22 galaxies without $T_e$ determinations, the correction of
the internal reddening is performed in the same way, except that the
theoretical intensity ratios are set to typical values of 
$I(\ha/\hb)=2.86$ (18 objects) or $I(\hg/\hb)=0.468$ (4 objects).

\begin{figure}
\centering
\includegraphics[angle=0,width=\columnwidth]{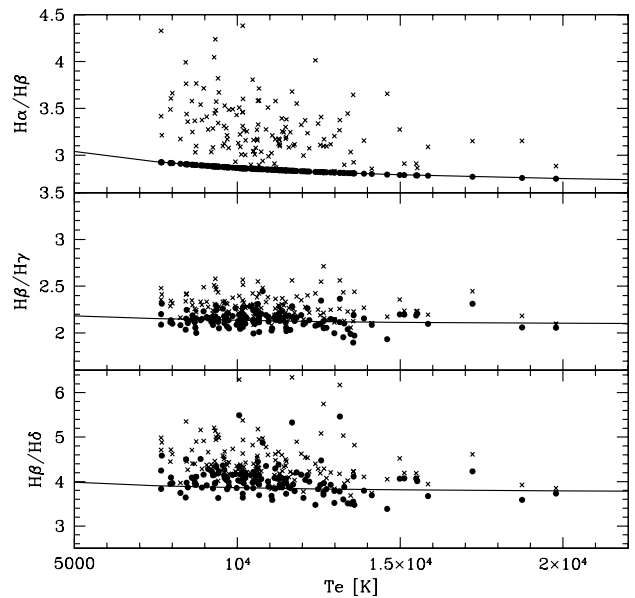}
\caption{
Intensity ratio of hydrogen emission lines as a function of electron 
temperature $T_e$, for WR galaxies in our sample that have been 
corrected for internal reddening. The observed ratios are plotted as 
crosses. The filled circles show the results that have been 
corrected for internal reddening using the method described in the 
text. The curves are for the theoretical values. 
[{\it See the electronic edition of the Journal for the color version 
of this figure.}]}
\label{fig:line_ratio}
\end{figure}

Fig.~\ref{fig:line_ratio} compares the corrected hydrogen emission
line ratios with the theoretical values for the galaxies that
have $T_e$ determinations and have been performed the correction 
based on the \ha/\hb\ intensity ratios.
The crosses and filled circles show the observed and corrected ratios
respectively. The lines are for the theoretical values. As can be
seen, the intensity ratios of hydrogen lines \ha, \hg\ and \hd\ 
relative to \hb\ are corrected substantially and change to be 
consistent with the theoretical values. 

Before going to the next section, it is necessary to find out the 
objects in our sample with clipped \ha\ or 
\oiii$\lambda\lambda4959,5007$ lines in their spectra, because the 
oxygen abundance of galaxies will be determined based on these lines. 
By comparing the observed and/or corrected emission line ratios with
the corresponding theoretical values, we find that 2 objects have been
clipped for \ha\ line, 1 for \oiii$\lambda\lambda$4959,5007, and 2 for 
both \ha\ and \oiii$\lambda\lambda$4959,5007 lines.

\subsection{Oxygen Abundance}

There are several empirical methods that have been proposed to 
estimate the oxygen abundance of galaxies. 
In this paper we adopt two of them: the classic $T_e$ method and the 
$N2$ method. In another paper (Shi \etal 2005), we have applied these 
methods to a sample of blue compact galaxies and discussed in detail 
the possible systematic differences among different methods. 
We present below a brief introduction to these methods and refer the 
reader to Shi \etal (2005) for a detailed description.

In the $T_e$ method, one first determines the electron temperature
$T_e$ and density $n_e$ of the gas using intensity ratios
\oiii$\lambda$4363/\oiii$\lambda\lambda4959,5007$ and
\sii$\lambda6716$/\sii$\lambda6731$ respectively, and then computes
the abundances of $\rm O^{+}$ and $\rm O^{+2}$ using the IRAF 
package {\tt NEBULAR} (Shaw \& Dufour 1995). In the computation, the 
$\rm O^{+}$ abundance is determined by the intensity of 
\oii$\lambda$3727 line or \oiii$\lambda\lambda$7320,7331, while the 
abundance of $\rm O^{+2}$ is determined by two \oiii~lines centered 
at 4959 and 5007\AA. Moreover, the fact that \heii$\lambda$4686 line 
is usually detected in the spectra of WR galaxies indicates that 
part of the oxygen in these systems must be in $\rm O^{+3}$ stage. 
Following Skillman \& Kennicutt (1993), we estimate the $\rm O^{+3}$ 
abundance by assuming $\rm O^{+3}/(O^++O^{+2}) \equiv He^{+2}/He^+$, 
where the helium abundances are determined using the 
\hei$\lambda6678$ and \heii$\lambda4686$ lines (Clegg 1987; Skillman 
\& Kennicutt 1993). 
Therefore, the total abundance of oxygen in WR galaxies is derived 
by $\rm O/H=(O^++O^{+2}+O^{+3})/H^+$ (Izotov \etal 1994; Skillman 
\& Kennicutt 1993).

As mentioned above, we have determined the electronic temperature 
$T_e$ for 152 out of 174 galaxies in our sample. We thus determine 
the oxygen abundance using the $T_e$ method for these galaxies. In 
this procedure, the $\rm O^{+}$ abundance is determined by the 
intensity of \oii$\lambda$3727 for 77 objects, while for the other 
75 with the non-observed \oii$\lambda$3727 line, the
\oiii$\lambda\lambda$7320,7331 lines are used.

For the remaining 22 galaxies without $T_e$ determinations, we
have to turn to a second method, because some of the 
emission lines mentioned above, which are necessary in the $T_e$
method, cannot be measured from their spectra due to the
limited wavelength coverage (e.g. in case of \oii$\lambda$3727), 
the low S/N (e.g. in case of \oiii$\lambda$4363) or clipping of
\oiii$\lambda\lambda$4959,5007. For these galaxies, we
adopt the so-called $N2$ method (van Zee \etal 1998), in which 
the oxygen abundance is estimated by
\begin{equation}
\zoh = 1.02\log [I(\nii)/I(\ha)]+9.36.
\end{equation}
Where $I(\nii)$ is $I(\nii)(\lambda6548+\lambda6583)$.
It should be pointed out that for 3 objects whose \ha\ lines
are clipped, the \ha\ fluxes are fixed to be 2.86$\times I(\hb)$.
We list in Table 1 the oxygen abundance for each object (Column 23) 
and the method adopted for it (Column 24).

As pointed out by Shi \etal (2005), the oxygen abundances derived 
from the $N2$ method are systematically larger by $\sim 0.2$dex 
than those from the $T_e$ method. 
We will show that this systematic difference does not affect our 
conclusion on the metallicity dependence of the IMF slope late.

\subsection{The WR and \hb\ emission}

In this paper we make comparisons with models for three quantities:
the EW of \hb\ emission, the EW of WR bump around 4650\AA\
(the blue bump), and the blue bump-to-\hb\ intensity ratio 
WRbump/\hb. As first pointed out by Copetti, Pastoriz, \& Dottori 
(1986), the EWs of hydrogen recombination lines, particularly
\hb, can be used as an accurate indicator of the age of the
population in the case of an instantaneous burst of star formation.
In addition, as pointed out by SV98, EW(\hb) as a function of
metallicity can be combined with WR lines/\hb\ intensity ratios
to determine the WR and O star populations. For the WR features,
we concentrate on the blue bump only because this feature is visible
in all the spectra (see \S2.2).

We have described in detail our method of measuring \hb\ flux in \S2.2
and the method of correcting for the effect of internal extinction in
\S3.1. The EW of \hb\ is measured in a similar way. However, The
situation becomes difficult when one trying to systematically measure 
the EW and intensity of WR bumps. As already mentioned in \S2.2, 
systematical measurements of WR bump strength usually suffer from 
large uncertainties in the continuum determinations and/or from the
contamination by the emission of non-WR stars such as
\feiii$\lambda4658$, \heii$\lambda4686$, \hei+\ariv$\lambda4711$, and
\ariv$\lambda4740$. We thus choose to manually measure the WR bump for
 all our WR galaxies. First, for each object, we use the best-fitting 
stellar spectrum, but not the observed spectrum, to determine the 
continuum. The continuum is then determined by fitting a cubic spline
 to seven wavelength channels (4020, 4510, 5313, 5500, 6080, 6630, 
7043\AA), which are thought to be fully free of nebular and stellar 
lines (Kong \etal 2002).
Next, we fit the non-WR emission lines mentioned above with Gaussian
functions and subtract them from the observed spectrum. We then 
measure the EW and the integrated flux of the blue bump with 
rest-frame wavelength limited to 4600-4750\AA.

\section{Comparison with model predictions}

In this section, we compare the observed WR features and \hb\ 
emission with predictions by evolutionary synthesis models of SV98. 
In the comparison, we divide our sample into four subsamples 
according to oxygen abundance and investigate whether/how the slope 
of the IMF depends on galaxy metallicity. We first briefly describe 
the models, and then present the results of the comparison. 

\subsection{Predictions by Evolutionary Synthesis Models}

The model predictions used in this paper are provided by SV98. These
authors constructed evolutionary synthesis models for young
starbursts, using stellar evolution models, theoretical stellar
spectra, and a complication of observed emission line strengths from
WR stars. They explicitly distinguish between the various WR subtypes
whose relative frequency is a strong function of metallicity. The
observational features predicted by their models allow a detailed
quantitative determination of the WR star population in a starburst
region from its integrated spectrum and provide a means of deriving
the burst properties (e.g. duration and age) and the slope of the IMF
of young starbursts. The model predictions should provide the most
reliable determinations to date, and can be used to investigate the
variation in stellar properties with metallicity.

In the models, the input parameters include the burst duration 
$\Delta t$, the metallicity $Z$, the IMF, and the total mass of stars
formed in the burst. As pointed out by SV98, the last parameter 
serves only as a normalization constant and is not relevant for the
quantities being studied. For the IMF, a power law 
$dN/dM \propto M^{-\alpha}$ between the upper and low cutoff masses, 
$M_{up}$ and $M_{low}$ respectively, is adopted. Following SV98, we 
set the cutoff masses to $M_{up}=120M_\sun$ and $M_{low}=0.8M_\sun$. 
For the slope of the IMF, we consider three different cases: a 
Salpeter slope with $\alpha=2.35$, a steeper slope with $\alpha=3.30$, 
and a flatter slope with $\alpha=1.00$. In each case, we assume an 
instantaneous burst with $\Delta t = 0$. 
In order to test the possible dependence on burst duration, we also 
consider in the case of Salpeter IMF four extended burst models with 
durations $\Delta t=$1, 2, 4 and 10 Myr. 
Finally, in order to study the dependence on metallicity, we consider 
four cases of metallicity: $Z$=0.001, 0.004, 0.008, and 0.02.

With each set of input parameters, the models predict many important
quantities which can be easily compared with observations. These
quantities are classified into three groups: population statistics 
(the number of stars with different spectral types), nebular 
quantities (the strength of nebular recombination lines: \ha, \hb, 
\heii$\lambda4471$, and \heii$\lambda4686$), and WR emission lines.
In this paper we study three of them: EW(\hb) (the EW of \hb\
emission line), EW(4650) (the EW of the blue WR bump), and 
WRbump/\hb\ (the ratio of blue WR bump intensity relative to \hb).

\subsection{Metallicity dependence of the IMF}

In order to study the metallicity dependence of the IMF and to make
comparisons with the model predictions described above, we divide our
sample into four subsamples according to oxygen abundance. Following
Guseva \etal (2000), we assume the solar oxygen abundance
[\zoh]$_{\sun}=8.93$ and use the oxygen abundance ranges listed below for
the sample division.
\begin{enumerate}
\item Z=0.001: $\zoh \le 7.93$;
\item Z=0.004: $7.93 < \zoh \le 8.43$;
\item Z=0.008: $8.43<\zoh\le8.63$;
\item Z=0.020: $\zoh>8.63$.
\end{enumerate}

\begin{figure*}
\centering
\includegraphics[angle=-0,width=\textwidth]{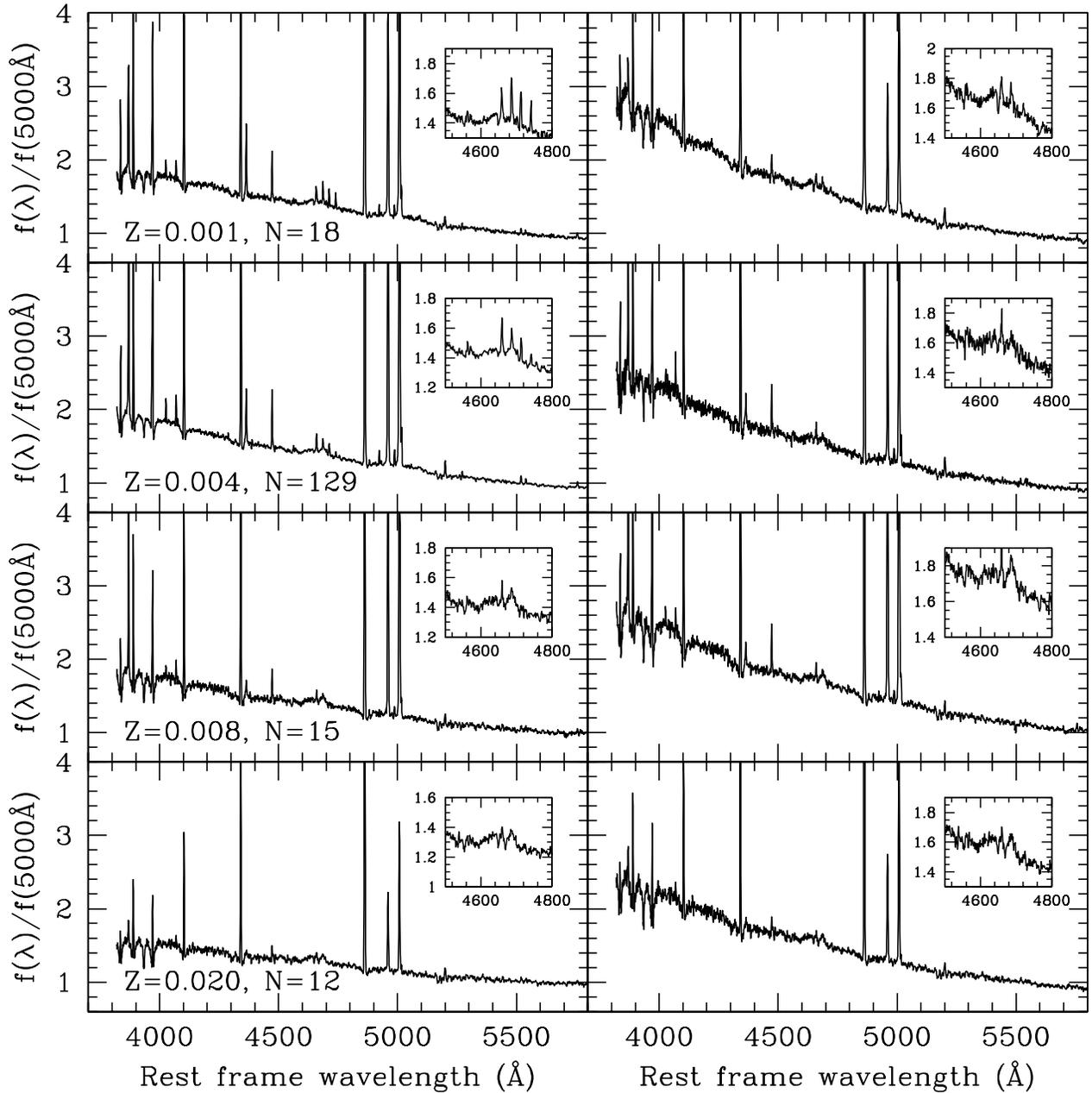}
\caption{Average spectrum of WR galaxies in different metallicity
subsamples. Panels from top to bottom correspond to different
metallicities, as indicated in the left-hand panels. The number of
objects in each subsample is also indicated. The panels at
left-hand are for the results obtained from the observed spectra that
have been corrected only for Galactic reddening, whereas the right-hand
panels are for those also corrected for the internal reddening.
The inset in each panels shows the WR bump around 4650\AA.}
\label{fig:spec}
\end{figure*}

Figure~\ref{fig:spec} shows the average spectrum of WR galaxies in
each of the four metallicity subsamples. In the calculation, each
observed spectrum is weighted by its S/N and normalized by the flux at
5000\AA. Panels from top to bottom correspond to different
metallicities, as indicated in the left-hand panels. The number of
objects included in each subsamples is also indicated. The panels at
left-hand show the results obtained using the observed spectra that
have been corrected only for Galactic reddening, while the right-hand
panels are for those also corrected for the internal reddening. The
inset in each panel shows the blue WR bump around 4650\AA. It can be
seen that, if the internal extinction was not taken into account, low
metallicity population shows harder continuum, as well as stronger
nebular emission lines, than does high metallicity population. After
the internal reddening being corrected for, the difference in the
shape of the continuum disappears and the continuum slope changes to
be very close to each other. This is reasonable because the dust
content in high metallicity galaxies is expected to be larger than
that in low metallicity ones. As a result, dust extinction will modify
the continuum more significantly for high metallicities. It is also
interesting to see that, the decrease of the strength of nebular
emission lines does exist for large metallicities, regardless of
whether the internal reddening is taken into account or not.

\begin{figure*}
\centering
\includegraphics[angle=-0,width=\textwidth]{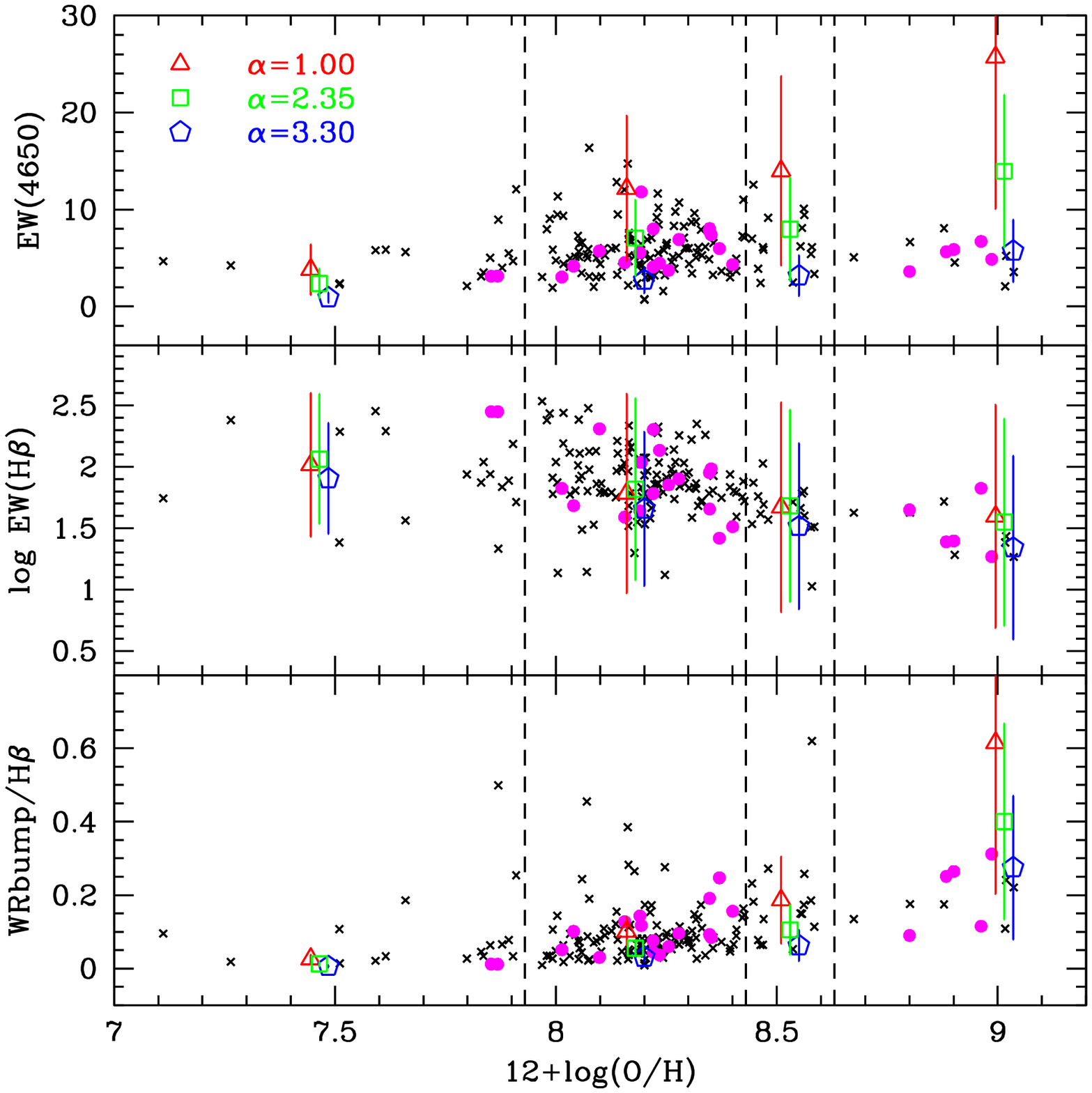}
\caption{
Equivalent width (EW) of WR bump around 4650\AA\ (top), EW of \hb\ 
emission line (median), and the intensity ratio of WR bump relative 
to \hb\ (bottom) are plotted as a function of oxygen abundance. 
The crosses are for all the WR galaxies in our sample, while the 
filled circles are for ``safe'' objects, defined as those with 
at least one of the five Petrosian radii smaller than the radius 
of the SDSS spectroscopic aperture ($1.5^{\prime\prime}$). 
The symbols with error bars show the average values with 1-$\sigma$ 
variance, which are predicted by models with different IMF slopes, 
as indicated in the top panel.
Dashed vertical lines denote the oxygen abundance ranges and the 
model predictions are plotted at the median of each range. For 
clarity, triangles, squares and pentagons are slightly shifted 
along the horizontal axis.
}
\label{fig:obs}
\end{figure*}

These observational results are shown more clearly in
Figure~\ref{fig:obs}, where we have plotted three quantities (from 
top to bottom: EW(4650), EW(\hb) and WRbump/\hb) as a function of 
oxygen abundance for all our WR galaxies (crosses).
As expected, the observed WRbump/\hb\ intensity ratio decreases with
decreasing metallicity. This result was first found by Arnault,
Kunth, \& Schild (1989) in an attempt to quantify the constrains on
the WR populations in starbursts, and were largely confirmed by later
studies. The result is believed to be due to the strong influence of
metallicity on stellar mass loss, which leads to a significant
decrease in the WR population at low metallicities (Maeder, Lequeux,
\& Azzopardi 1980). Moreover, the rapid decrease of EW(\hb) with
increasing oxygen abundance is clearly seen from this figure. In 
contrast, we see no significant correlation between EW(4650) and
metallicity. The increase of WRbump/\hb\ with metallicity is thus
in large part due to the rapid decrease of \hb\ for large
metallicities, which is well consistent with theoretical expectations
(see below; also see Arnault \etal 1989).

Figure~\ref{fig:obs} also compares these quantities to predictions by 
the three instaneous bust models with different IMFs ($\alpha=$1.00, 
2.35, 3.30) and metallicities ($Z=$0.001,0.004, 0.008, 0.020). 
We find that models with flatter IMFs tend to predict larger values 
for all these quantities and at all metallicities. This is reasonable 
because a flat IMF implies a relatively large population of massive 
stars, leading to strong WR and nebular emission. In particular, such 
metallicity dependence is much stronger in EW(4650) and WRbump/\hb\ 
than in EW(\hb). We also see that the observed trends in the three 
quantities with oxygen abundance can generally be reproduced by all 
the models. However, the strength of the predicted trends is different 
for models with different IMFs. For example, a flatter IMF will lead 
to more rapid increase of WRbump/\hb\ with metallicity. As a result, 
the difference between model predictions becomes much more remarkable 
at high metallicities, reaching a factor of $\sim5$ in EW(4650) and 
$\sim3$ in WRbump/\hb\ on average at $Z>Z_\odot$. 
It is thus inevitably to see that a model with a fixed IMF can not 
simultaneously match the observations at all metallicities. At the
highest metallicity ($Z=0.02$), only the model with the steepest IMF
($\alpha=3.30$) can well match the observed EW(4650) and WRbump/\hb. 
In case of $Z=0.001$, however, just the reverse is likely to be true. 
In another word, Fig.~\ref{fig:obs} indicates that a dependence of 
the IMF slope on galaxy metallicity is required to explain the 
observed trends in WR and \hb\ emission.

\begin{figure*}
\centering
\includegraphics[angle=-0,width=\textwidth]{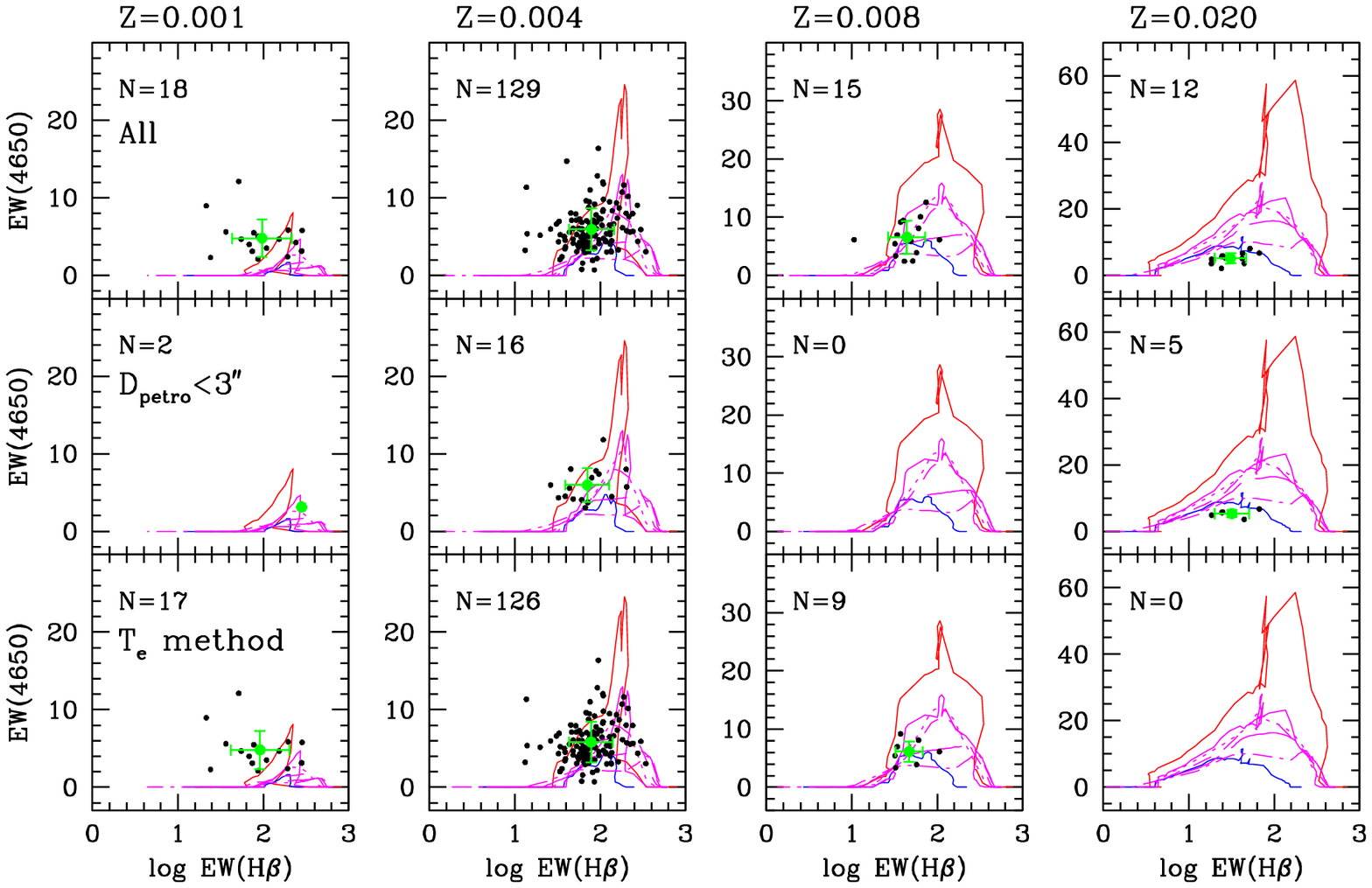}
\vspace{-6.0cm}
\caption{
Distribution in the space of EW(4650) versus EW(\hb) for WR galaxies 
in different metallicity subsamples (black points). Panels from left 
to right correspond to different metallicities, as indicated above 
the top panels. The top panels are for the whole sample. The median 
panels are for objects with at least one of the five Petrosian radii 
smaller than the radius of the SDSS spectroscopic aperture.
The bottom panels are for objects with their oxygen abundances
determined from $T_e$ method. 
In each panel, the average location of the objects is plotted as a 
green point and the $1-\sigma$ dispersion between them is indicated 
by the error bars. The solid lines show the predictions by the three 
instaneous burst models, with red, magenta and blue colours, 
respectively, corresponding to models with IMF slope $\alpha$=1.00, 
2.35 and 3.30.
The magenta lines in other styles are for the four extended burst
models with fixed IMF slope ($\alpha=2.35$) but various burst 
durations: $\Delta t$= 1 (dotted), 2 (dashed), 4 (long dashed), 
10 Myr (dotted-dashed).}
\label{fig:bump_hb}
\end{figure*}

\begin{figure*}
\centering
\includegraphics[angle=-0,width=\textwidth]{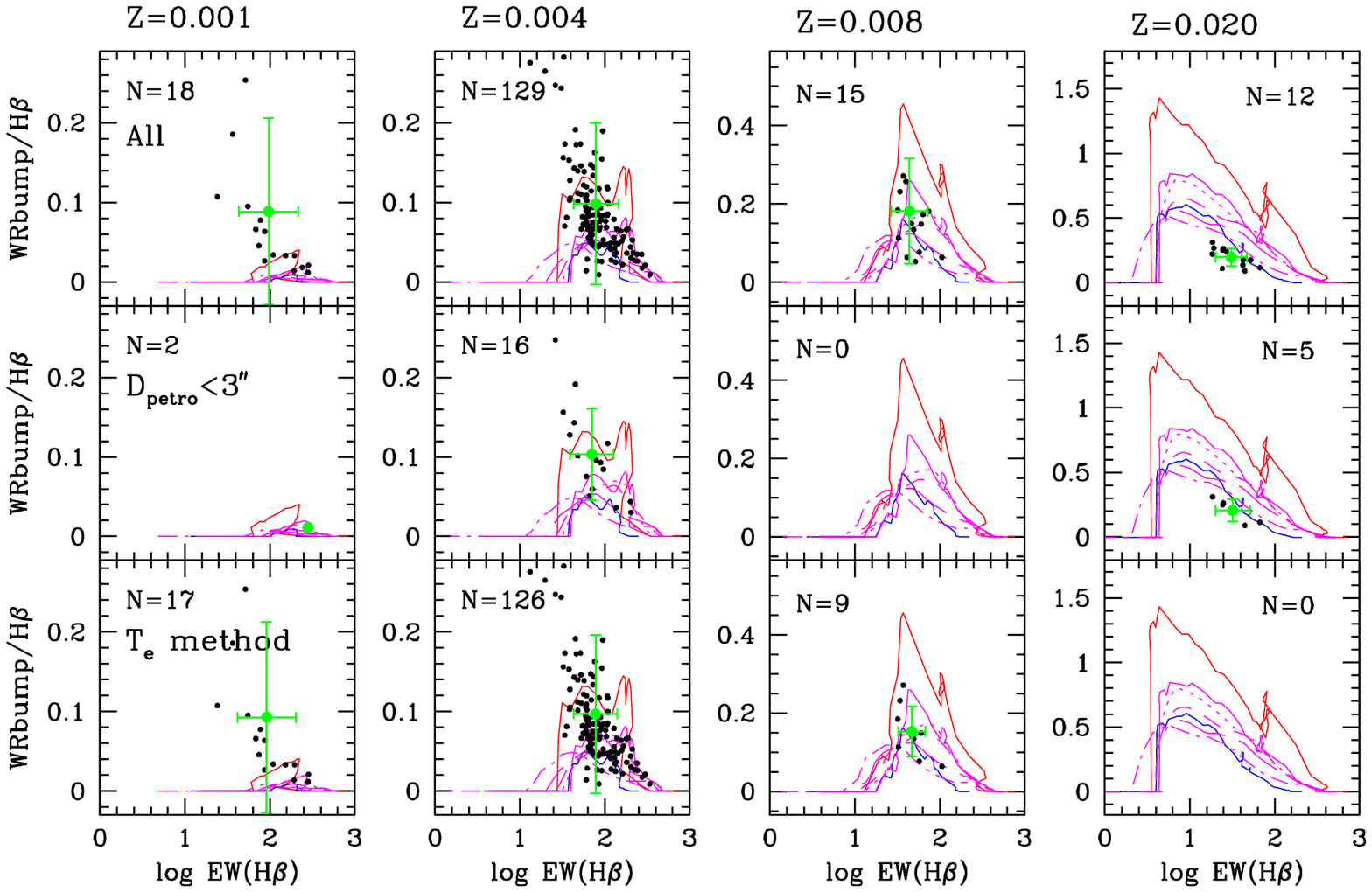}
\vspace{-6.0cm}
\caption{
Distribution in the space of WRbump/\hb\ versus EW(\hb), for WR 
galaxies in different metallicity subsamples. The symbols and the
lines are the same as in Figure~\ref{fig:bump_hb}.
}
\label{fig:bumphb_hb}
\end{figure*}

Figures~\ref{fig:bump_hb} and ~\ref{fig:bumphb_hb} (black points in 
the top panels) show the distribution in spaces of EW(4650) versus 
EW(\hb) and WRbump/\hb\ versus EW(\hb) respectively, for WR galaxies 
in different metallicity subsamples (as indicated above each panel). 
In each panel, the average location of the objects is plotted as a 
green point and the $1\sigma$ dispersion between them is indicated by 
the error bars. The solid lines show the predictions by the three 
instaneous burst models, with red, magenta and blue colours 
respectively corresponding to models with IMF slope $\alpha$=1.00, 
2.35 and 3.30. The magenta lines in other styles are for the four 
extended burst models with fixed IMF slope ($\alpha=2.35$), but 
various burst durations: $\Delta t$= 1 (dotted), 2 (dashed), 4 (long
dashed), 10 Myr (dotted-dashed). A clear dependence of the IMF slope 
on metallicity can be seen from the two figures, in that WR galaxies 
with higher metallicities exhibit steeper IMF. As can be seen, most 
objects with $Z=0.02$ are located well below the $\alpha=3.30$ curve, 
implying that these galaxies have IMFs similar to or even steeper 
than the steepest IMF that we have probed. In contrast, the objects 
of $Z=0.001$ are dominantly above the $\alpha=1.00$ curve, and so 
should be associated with very flat IMFs. 
At intermediate metallicities, the scatter in the observational 
distributions is very large, but the observations are still on 
average consistent with the Salpeter IMFs, the ones having the 
intermediate slope.

\section{Sources of Systematic Error}

In the previous section, we have seen a clear dependence of the IMF 
slope on metallicity by comparing our measurements of WR features and 
\hb\ with the model predictions provided by SV98. However, there are 
possible uncertainties in this result, caused by a few sources of 
systematic error either in the models or in the observational 
measurements. We discuss these in detail in this section.

\subsection{Aperture effects}

Each of the SDSS spectra was obtained using a 3 arcsecond diameter 
fibre that had been positioned as close as possible to the centre 
of the object being studied. One may thus wonder to what extent our 
measurements of EW(4650), EW(\hb) and WRbump/\hb\ may be biased 
because the measurements are not accurately reflecting the stellar and 
gaseous content of the whole starburst region. For example, because of 
aperture effects, both EW(4650) and EW(\hb) may be somewhat 
underestimated. As a result, the locations of galaxies on each panel 
of Figure~\ref{fig:bump_hb} may be biased toward lower-left corner and 
thus mismatch the theoretical curves.

In order to address the effect of aperture bias, we define a subsample 
of ``safe'' objects as those fully enclosed within the spectroscopic 
aperture. We use the Petrosian radii in the five SDSS photometric 
bands provided by the SDSS pipeline to quantify the angular size and 
select ``safe'' objects by requiring that at least one of the five 
radii is smaller than the radius of the aperture. This results in a 
subsample of 23 WR galaxies whose spectra are not suffering from the 
aperture effect. Though much smaller in size, this subsample will 
provide a critical test on the results obtained from the whole sample.

In Figure~\ref{fig:obs}, the ``safe'' objects are highlighted as 
filled circles. In case of EW(4650), it is likely that there is no 
significant bias in the whole sample relative to the ``safe'' sample. 
In contrast, a fraction of the whole sample shows much smaller EW(\hb) 
compared to the ``safe'' sample. We thus conclude that all emission of 
WR stars is likely inside the fibre, whereas aperture bias could well 
be a problem for \hb. This can be understood as \hb\ emission is more
extended. As a result, WRbump/\hb\ intensity ratio is overestimated 
to varying degrees for some of the objects.

The relations between the three quantities obtained from the "safe" 
sample are plotted in the second rows of panels of 
Figures~\ref{fig:bump_hb} and ~\ref{fig:bumphb_hb}. The symbols and 
the lines are the same as in top panels and the number of objects 
included in each metallicity subsample is also indicated. It is 
encouraging to see that the metallicity dependence of the IMF slope 
still exists.  We are thus convinced that the effects of aperture do 
not change our main results.

\subsection{Oxygen abundance}

In our sample, the oxygen abundance of 152 objects is determined from
 the $T_e$ method, while for the others we have adopted the $N2$ 
method. As mentioned above, the measurements based on the $N2$ method 
are systematically larger than those based on the $T_e$ method, by a 
factor of $\sim0.2$ dex. 
Moreover, for 27 galaxies with \oiii$\lambda4959/$\hb $<$ 0.7, the EW 
of \hb\ is generally low and the \oiii$\lambda$4363 line is weak, 
hence their oxygen abundances might be underestimated (Izotov et al. 
2004). It is thus necessary to test how reliable the methods are and 
whether and how they affect our main results.

\begin{figure}
\centering
\includegraphics[angle=-0,width=\columnwidth]{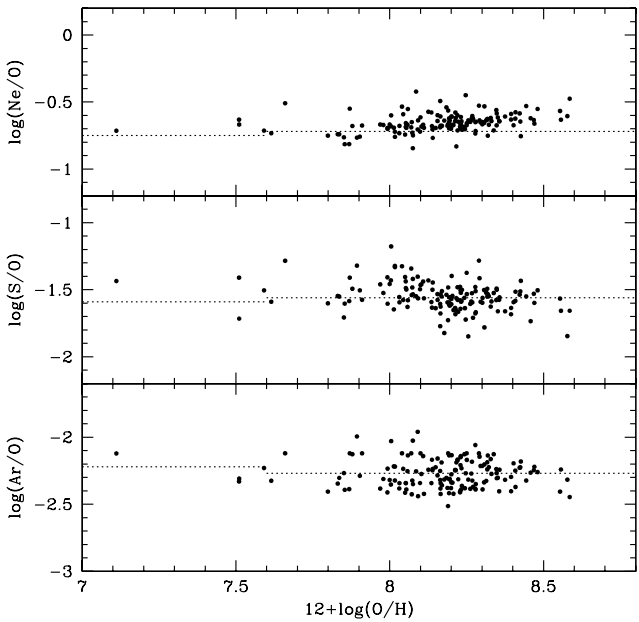}
\caption{The Ne/O, S/O, and Ar/O abundance ratios of 152 WR galaxies
whose oxygen abundance are determined from the $T_e$ method.}
\label{fig:check_oxygen}
\end{figure}

For $T_e$ method, it is possible to address this problem in a 
statistical way by comparing the Ne/O, S/O and Ar/O abundance ratios 
and the average values found for galaxies with different oxygen 
abundances. If these ratios are not consistent with the average 
values, the metallicity might be somewhat underestimated (e.g. Izotov 
et al. 2004). 
Adopting a two-zone photoionized \hii\ region model described by 
Izotov, Thuan, \& Lipovetsky (1994), for each of the 152 objects whose 
oxygen abundances are determined from the $T_e$ method, we calculate 
its $\rm Ne^{++}/O^{++}$, $\rm S^+$ and 
$\rm S^{++}$ using expressions from Pagel et al. (1992), 
and $\rm Ar^{++},~Ar^{+++}$ using IRAF {\tt NEBULAR} package (see Shaw 
\& Dufour 1995). We then calculate their Ne/O, S/O and Ar/O abundance 
ratios, using the ionization correction factors (ICFs) presented by 
Izotov, Thuan, \& Lipovetsky (1994). The results are shown in
Figure~\ref{fig:check_oxygen}. 
The dotted lines are for the average values found by Izotov \& Thuan 
(1999). It is seen that the abundance ratios are well consistent with 
the average values at all oxygen abundances.

Figure~\ref{fig:check_oxygen} demonstrates that our calculation of 
oxygen abundance based on the $T_e$ method is reliable and robust. 
We thus select the 152 objects with oxygen abundance determined from 
the $T_e$ method to construct a second "safe" sample, in order to 
exclude any possible biases in our results caused by the $N2$ method. 
Similarly, we plot the results for this "safe" sample in the third 
rows of panels of Figures~\ref{fig:bump_hb} and ~\ref{fig:bumphb_hb} 
and obtain a conclusion that our results are not significantly biased 
by the uncertainties in oxygen abundance determinations.
Note that there are no objects included in the $Z=0.020$ subsample. 
This is due to the fact that the application of $T_e$ method requires 
an accurate determination of the \oiii$\lambda$4363 intensity, but in
general this line is very weak or invisible in the spectra of 
high-metallicity galaxies.
Nevertheless, the rest three panels still show marked increase of IMF 
slope with increasing metallicity, and thus our conclusion about the 
metallicity dependence keeps unchanged.

\subsection{Spectral modelling}

In this paper, the EWs are measured adopting the best-fitting stellar 
continuum. However, the contribution of the ionized gas emission could 
be high in \hii\ regions, especially for those with high EW(\hb). 
The most recent paper by Guseva \etal (2006), where the ionized gas
emission is considered in the fitting procedure, shows that the flux
fraction of the gaseous continuum near \hb\ can reach $\sim 50\%$
in \hii\ regions with EW(\hb)$\sim 400$\AA.
Therefore, the ionized gas continuum template should be included in
the spectral modelling, otherwise the EWs will be underestimated.
Since our templates are constructed from simple stellar
populations, our method currently can not be able to include the gas
continuum in the fitting. Moreover, the absolute majority of our
sample are low redshift objects in which the Balmer continuum is not
within the SDSS wavelength coverage. It is thus very difficult to 
break the degeneracy between gaseous continuum and very young stellar 
populations only using the optical spectra without this feature.

In addition, the measured EWs are compared with the model predictions 
for instantaneous or short bursts, and thus the EWs should be obtained 
adopting the continuum level only of the youngest template that is 
most likely associated with the WR stars. Therefore, adopting the 
best-fitting stellar continuum will also lead to underestimated EWs. 
If the ratio of the youngest template continuum to the total one 
systematically varies at different metallicities, then the results 
shown in Figures~\ref{fig:bump_hb} and ~\ref{fig:bumphb_hb} may also 
be biased. We have checked the relative contributions to the 
best-fitting spectra by the first two templates, which roughly 
represent the spectra of O and B stars respectively (see \S2.3), and 
found no tendency of these contributions with metallicity. Therefore, 
although the EWs have somewhat been underestimated, the overall 
correlation between the measured EWs and the metallicities can still 
exist and thus the trend of the IMF slope with metallicity can still 
be true.

\subsection{Synthesis models}

In this paper, the theoretical predictions are made using the 
evolutionary synthesis models of SV98. It is worthy noticing that 
their models are based on the evolution of non-rotating stars. Stellar 
rotation is expected to predict longer WR stage, lower WR star masses, 
and thus larger WR populations. As mentioned above, the strong 
influence of metallicity on stellar mass loss leads to a significant 
decrease in the WR population at low metallicities, resulting in rapid 
decrease of WRbump/\hb\ intensity ratio at these metallicities. 
Therefore, if rotating stars are included in the models, both EW(4650)
 and WRbump/\hb\ would be larger than that predicted by the 
current models. As a result, the predicted curves in the 
low-metallicity panels of Figures~\ref{fig:bump_hb} and 
~\ref{fig:bumphb_hb} would go upward, and the observed decrease of the 
IMF slope at low metallicities would be less pronounced. New models of 
rotating stars are thus needed to be worked out, in order to provide 
more reliable predictions, especially for low-metallicity populations.

Moreover, SV98 have compiled average line fluxes for WR stars in the 
LMC (for models with $Z<Z_\odot$) or the Galaxy (for models with 
$Z>Z_\odot$), which represent the standard calibration for WR 
populations in external galaxies. There has, however, been a recent 
study of the luminosities of WR emission lines, reaching the 
indication that WR stars at SMC metallicities possess lower optical 
line luminosities than those in the LMC (Crowther \& Hadfield 2006). 
This could be another source of systematic error which may also leads 
to underpredicted WR populations. 

It is thus important to answer the questions that to what extent the 
two effects described above would change the predictions at low 
metallicities and which effect should be more dominant in such bias. 
However, at the moment, there are no published evolutionary synthesis 
models with rotating stars and more recent WR line luminosities. 
Nevertheless, one can still expect that these effects do not 
significantly change the results at high metallicities, for which the 
observed metallicity dependence of the IMF slope can still exist.

\section{Summary}

In this paper, we have analyzed the WR features and \hb\ emission of 
174 WR galaxies drawn from the Sloan Digital Sky Survey. By comparing 
these observed quantities with predictions by evolutionary synthesis 
models of Schaerer \& Vacca (1998), we investigate whether and how the 
stellar initial mass function depends on metallicity. Our sample is
constructed using a two-step method, in which we first select 
candidates as the star-forming galaxies with evident \he\ emission 
lines and then select WR galaxies by visually examine the spectra of 
the candidates.
Although this method has inevitably exclude a number of objects 
(mainly with low S/N), it is still work well for our purpose and gives 
rise to a large enough, high quality sample of WR galaxies. We have 
carefully examine the possible sources of systematic error either in 
models or in our observational measurements and shown that these 
sources do not change our main results.

The conclusions of this paper can be summarized as follows.
\begin{enumerate}
\item The continuum of the average spectrum of WR galaxies shows no
  significant dependence on metallicity. In contrast, low-metallicity
  galaxies exhibit in their spectra much stronger nebular emission
  than high-metallicity galaxies.
\item The observed equivalent width (EW) of \hb\ decreases with
  increasing metallicity, whereas there is no significant trend
  in EW of WR blue bump around 4650\AA\ with metallicity. 
\item We have found a clear dependence of the IMF slope on metallicity
  in that the IMF slope increases with increasing metallicity. Such 
  dependence has been probed by many previous studies, but it is much 
  more pronounced in this paper due to the improvement in the 
  observational data.
\end{enumerate}

It should be pointed out that there is still room for improvement of 
this study. First, the method of sample selection could be developed 
as to fully take advantage of the large number of SDSS galaxies. 
Second, more accurate measurements of WR lines and new models
incorporating stellar rotation are needed to refine the results,
especially for low metallicities.
Third, the aperture effects should be corrected in a statistically 
robust way incorporating the stellar contents obtained in our 
starlight fitting procedure and the method of spectral modelling 
could be improved.
Finally, more quantities such as the relative numbers of WR stars with
different subtypes could be included in the comparison to give more
subtle and convincing results. These will be the aims of subsequent 
papers.

\acknowledgments 
We thank the anonymous referee for useful and constructive comments 
that resulted in a significant improvement of this paper.
We are grateful to Dr. Daniel Schaerer for providing the model
predictions and for his helpful comments. We thank the SDSS teams for
making their data publicly available, and Dr. Yuri I. Izotov for 
helpful discussion.
This work is supported by the Chinese National Science Foundation
through NSF10573014, NSF10633020, and by the Bairen Project of the 
Chinese Academy of Sciences.

Funding for the SDSS and SDSS-II has been provided by the Alfred P. 
Sloan Foundation, the Participating Institutions, the National Science
Foundation, the U.S. Department of Energy, the National Aeronautics 
and Space Administration, the Japanese Monbukagakusho, the Max Planck 
Society, and the Higher Education Funding Council for England. 
The SDSS is managed by the Astrophysical Research Consortium for the 
Participating Institutions. The Participating Institutions are the 
American Museum of Natural History, Astrophysical Institute Potsdam, 
University of Basel, Cambridge University, Case Western Reserve 
University, University of Chicago, Drexel University, Fermilab, the
Institute for Advanced Study, the Japan Participation Group, Johns 
Hopkins University, the Joint Institute for Nuclear Astrophysics, the 
Kavli Institute for Particle Astrophysics and Cosmology, the Korean 
Scientist Group, the Chinese Academy of Sciences (LAMOST), Los Alamos 
National Laboratory, the Max-Planck-Institute for Astronomy (MPIA), 
the Max-Planck-Institute for Astrophysics (MPA), New Mexico State 
University, Ohio State University, University of Pittsburgh, 
University of Portsmouth, Princeton University, the United States 
Naval Observatory, and the University of Washington.

\end{document}